\documentclass[preprint,12pt]{elsarticle}

\usepackage{booktabs}
\usepackage{array}
\usepackage{url}
\usepackage{xurl}
\usepackage{float}
\usepackage{microtype}
\usepackage[hidelinks,breaklinks=true]{hyperref}
\usepackage{amsmath}
\usepackage[T1]{fontenc}
\usepackage[utf8]{inputenc}
\journal{Information and Software Technology}

\begin{document}

\begin{frontmatter}

\title{When Retrieval Hurts Code Completion: A Diagnostic Study of Stale Repository Context}

\author[ca]{Haojun Weng\corref{cor1}\fnref{cofirst}}
\ead{whjwitness1019@gmail.com}

\author[bj]{Qianqian Yang\fnref{cofirst}}
\ead{546987836@qq.com}

\author[ca]{Hao Fu}
\ead{koi.helloai@gmail.com}

\author[ca]{Haobin Pan}
\ead{bayron.p27@gmail.com}

\author[bj]{Xinwei Lv}
\ead{18700910021@163.com}

\address[ca]{Independent Researcher, California, USA}
\address[bj]{Independent Researcher, Beijing, China}

\cortext[cor1]{Corresponding author. ORCID: 0009-0006-7306-4816.}
\fntext[cofirst]{These authors contributed equally as co-first authors.}

\begin{abstract}
\textbf{Context:} Retrieval-augmented code generation relies on cross-file repository context, but retrieved snippets may come from obsolete project states.
\textbf{Objectives:} We study whether temporally stale repository snippets act as harmless noise or actively induce current-state-incompatible code.
\textbf{Methods:} We conduct a controlled diagnostic study on a curated 17-sample set of production-helper signature changes from five Python repositories. For each sample, we compare current-only, stale-only, no-retrieval, and mixed current/stale retrieval conditions under prompts that hide commit freshness and expected current signatures.
\textbf{Results:} Under neutralized prompts, stale-only retrieval induces stale helper references on 15/17 \texttt{Qwen2.5-Coder-7B-Instruct} samples and 13/17 \texttt{gpt-4.1-mini} samples, corresponding to 88.2 and 76.5 percentage-point increases over current-only retrieval. No retrieval produces zero stale references but only 1/17 passing completions. The two models share 75.0\% Jaccard overlap among stale-triggering samples, and mixed conditions show that adding valid current evidence largely rescues stale-only failures.
\textbf{Conclusion:} Temporal validity of retrieved repository context is a distinct diagnostic variable for Code RAG robustness: stale context can actively bias models toward obsolete repository state rather than merely removing useful evidence.
\end{abstract}


\end{frontmatter}

\section{Introduction}

Modern code assistants increasingly rely on repository context. A model asked
to complete a function rarely has enough information in the current file alone:
it may need helper signatures, local conventions, wrappers, or cross-file
dependencies. Retrieval-augmented code generation addresses this by supplying
snippets from the surrounding repository or from external code knowledge bases.
This can turn an otherwise underspecified completion task into one the model can
solve. Yet the same mechanism also creates a new failure channel. If the
retrieved snippet is plausible but temporally stale, the model may ground on a
repository state that no longer exists.

This risk is realistic in active repositories. Helper functions are renamed,
arguments are added or removed, wrapper signatures drift, and project-internal
utilities change across commits. A retriever, cache, index, or external code
search system may return a snippet from an older state of the repository. Such
a snippet is not obviously irrelevant: it comes from the same project, often
uses the same helper name, and may look locally compatible with the current
task. The question is whether code models treat this stale evidence as harmless
noise, or whether it can actively steer them toward current-state-incompatible
code.

Prior work has studied several adjacent problems. API-evolution and
version-aware code-generation studies show that models can lag behind changing
software ecosystems \cite{llmsLagBehind2026,rustEvo2025}. Repository-level
code RAG systems and benchmarks study how to retrieve cross-file context
for project-level completion \cite{repoCoder2023,codeRag2025,repoBench2023,crossCodeEval2023}.
Context-filtering and mitigation methods show that retrieved chunks can be
helpful, neutral, or harmful \cite{impactDrivenContext2025,repoShapley2026,repoformer2024}. Hallucination and API-misuse work
studies incorrect code and invalid API use \cite{codeHalu2024,apiMisuse2025}. To our knowledge, the closest related work in recent Code RAG, selective
retrieval, software-evolution, and benchmark generation does not systematically
control whether retrieved repository context comes from an old commit or the
current repository state while measuring stale-reference outcomes (see
Supplementary Material S1 for the related-work audit).

We present a controlled diagnostic study of stale repository context in
retrieval-augmented code completion. The study manipulates the temporal
validity of retrieved cross-file context while holding the local task and
oracle fixed. For each sample, the model sees a local wrapper task and one of
five retrieval conditions: current context only, stale context only, no
retrieval, current-then-stale mixed retrieval, or stale-then-current mixed
retrieval. The prompt deliberately avoids explicit target-state anchors: it
does not tell the model which snippet is current, does not reveal commit
freshness, and does not name the expected current helper signature. This design
follows earlier pilot evidence that strong models can self-rescue when the task
itself explicitly names the intended API or library state.

Our dataset contains 17 curated production-helper signature changes mined from
five Python repositories: \texttt{click}, \texttt{flask}, \texttt{httpx}, \texttt{requests}, and \texttt{rich}.
Each sample pairs a stale helper signature from a parent commit with the
current helper signature from the child commit, and uses static call-pattern
oracles to classify model outputs as current-state matches, stale references,
or fail-no-match outputs. We evaluate \texttt{Qwen2.5-Coder-7B-Instruct} as an
open-source code-specialist model and \texttt{gpt-4.1-mini} as a proprietary
general-purpose model.

The results show a strong stale-context effect. Current-only retrieval
eliminates stale references in both models, while stale-only retrieval
frequently induces calls that match obsolete helper signatures. The
no-retrieval baseline is also informative: without repository context, models
mostly fail to match either oracle rather than producing stale-state calls.
Thus stale retrieval is not merely equivalent to missing context. It redirects
the error pattern from generic inability to complete the wrapper toward
historical repository-state grounding. A sensitivity run on the pre-quality-gate
sample set confirms that the main effect is not an artifact of Qwen-based
filtering.

We also find that valid current evidence presence is more important than
retrieval rank order in the current setup. When both current and stale snippets
are retrieved, stale references drop sharply relative to stale-only retrieval.
Both models show 0.0 percentage-point aggregate rank-order deltas under the
neutralized evaluation. The safe conclusion is that presence or absence of
valid current evidence is the dominant mechanism in this diagnostic setting.

This paper makes four contributions:

\begin{enumerate}
\item We introduce a controlled diagnostic setup for repository-level Code RAG that isolates temporal validity of retrieved cross-file context.
\item We provide cross-model evidence on 17 production-helper samples from five Python repositories, showing large stale-reference effects in both \texttt{Qwen2.5-Coder-7B-Instruct} and \texttt{gpt-4.1-mini}.
\item We identify a mechanism-level pattern: stale retrieval actively induces stale-state references, no retrieval produces a different fail-no-match pattern, and valid current evidence largely rescues stale-only failures.
\item We release a prompt-rendering and leakage-audit artifact, including opaque context identifiers, forbidden-token checks over model-visible fields, prompt dumps, and excluded preliminary runs for audit transparency.
\end{enumerate}

We position this as a diagnostic study, not a mitigation method. The goal is
to characterize a failure mode that future retrievers, index freshness checks,
and context filters can address.

The rest of the paper describes the mining and evaluation protocol, reports the
controlled-condition results, and discusses limitations including sample size,
signature-change scope, one-parent-commit drift, and static regex oracles.

\section{Related Work}

\subsection{API Evolution and Version-Aware Code Generation}

Recent work shows that code generation models can lag behind evolving software
ecosystems. A study of Python API evolution documents that model outputs often
reflect outdated APIs and that additional structured API information can
improve generation under changing libraries \cite{llmsLagBehind2026}. RustEvo\textsuperscript{2}
extends the version-aware evaluation perspective to the Rust ecosystem \cite{rustEvo2025}. EVOR similarly treats retrieval as an evolving process:
instead of relying on a fixed knowledge source, it updates retrieval queries
and knowledge bases to support code generation in settings where external
knowledge changes over time \cite{evor2024}.
These works establish software evolution as an important source of code-model
failure, but their main variables are API/library evolution and knowledge-base
adaptation. Our study asks a narrower retrieval-level question: when repository
context is retrieved for code completion, what happens if that context comes
from an old repository state rather than the current one? This is orthogonal
to EVOR's focus: EVOR updates the retrieval pipeline and knowledge base, while
we hold the retrieval source family fixed and vary the historical snapshot
selected as context.

\subsection{Repository-Level Code RAG}

Repository-level code completion has motivated a family of methods and
benchmarks that retrieve cross-file context beyond the current editing buffer.
RepoCoder introduced iterative retrieval and generation for repository-level
completion, using generated code to refine subsequent retrieval queries
\cite{repoCoder2023}. RepoBench and CrossCodeEval evaluate repository
or cross-file completion settings where models must use project-level context
rather than isolated function bodies \cite{repoBench2023,crossCodeEval2023}. More recent systems such as CodeRAG improve
repository-level retrieval through query construction, multi-path retrieval,
and reranking over structured code knowledge \cite{codeRag2025}.
CodePlan studies repository-level coding through planning over code edits and
build/type-check feedback, making it a close neighbor for temporal code-change
settings \cite{codePlan2023}.
SWE-Bench++ scales repository-level software-engineering benchmarks by
turning live pull requests into executable tasks with test-time repository
oracles \cite{sweBenchPlusPlus2025}. Its before/after state construction is
adjacent to our temporal framing, but the manipulated variable is different:
SWE-Bench++ varies the benchmark task and repository-state oracle, whereas our
study holds the local task and oracle fixed while varying the commit of the
retrieved snippet.

This line of work demonstrates that repository context can be useful and that
retrieval design matters. These systems generally evaluate retrieval quality,
completion quality, planning, or benchmark performance under a given repository
snapshot.

\subsection{Context Filtering and Mitigation}

A second line of work studies when retrieved context should be used, filtered,
or constrained. Impact-driven context filtering labels retrieved cross-file
chunks as positive, neutral, or negative for completion and shows that harmful
context can degrade code generation \cite{impactDrivenContext2025}. Repoformer
selectively decides whether to retrieve for repository-level completion \cite{repoformer2024}, while RepoShapley uses Shapley-style attribution to model
the contribution of retrieved chunks and filter context accordingly \cite{repoShapley2026}. MARIN addresses API hallucination from a different
mitigation angle: it uses hierarchical dependency awareness to reduce invalid
API use in project-specific code generation \cite{marin2025}.

These works are important because they make clear that retrieval is not
uniformly helpful. They also form a crowded area that our paper deliberately
does not claim as new: we do not propose a new context filter, reranker,
dependency-aware decoder, or mitigation method.

\subsection{Hallucination and Robustness in Code Generation}

Code hallucination and API misuse have been studied from benchmark, taxonomy,
and mitigation perspectives. CodeHalu investigates code hallucinations through
execution-based verification \cite{codeHalu2024}. Package hallucination
work studies nonexistent package suggestions and their supply-chain risks,
which is adjacent to but distinct from real repository-state mismatch \cite{packageHallucinations2024}. API misuse work categorizes and mitigates incorrect API use
within generated code, including misuse of existing APIs \cite{apiMisuse2025}. These studies provide the broader robustness motivation for
our stale-reference metric: generated code can be wrong not only because a model
lacks knowledge, but because it grounds on an incompatible external signal.

\subsection{Positioning of This Study}

Across recent work, we found substantial coverage of API evolution,
repository-level retrieval, context filtering, and code hallucination. We
therefore avoid broad claims such as "stale context is unstudied" or "harmful
retrieval is unexplored." To our knowledge, the narrower gap is
retrieval-temporal: the closest related studies reviewed here do not treat old-vs-current
retrieved repository context as a controlled variable and measure
stale-reference behavior under otherwise matched code-completion tasks (see
Supplementary Material S1 for the related-work audit).

We fill this gap with a controlled diagnostic study rather than a new retrieval
method. The retrieved snippets in our study are real code from the same project
lineage rather than nonexistent packages or arbitrary incorrect API
suggestions. The manipulated variable is retrieval-time temporal validity:
current-context-only, stale-context-only, no-retrieval, and mixed current/stale
retrieval. This framing complements prior work: API evolution studies motivate
the temporal risk, repository-level RAG work motivates the retrieval setting,
context-filtering work motivates why retrieval can hurt, and hallucination work
motivates the stale-reference outcome. The contribution is to isolate temporal
validity of retrieved repository context as the diagnostic variable.

\section{Method}

\subsection{Study Type}

We conduct a controlled diagnostic study of repository-level code RAG. The
study isolates the temporal validity of retrieved cross-file context as the
manipulated variable:

\begin{center}
\begin{tabular}{ll}
\toprule
Role & Retrieval condition \\
\midrule
Control & \texttt{current\_context\_only} \\
Treatment & \texttt{stale\_context\_only} \\
Baseline & \texttt{no\_retrieval} \\
Robustness & \texttt{mixed\_current\_top1\_stale\_top2} \\
Robustness & \texttt{mixed\_stale\_top1\_current\_top2} \\
\bottomrule
\end{tabular}
\end{center}

The study is diagnostic rather than a mitigation method. We do not train a
model, propose a new retriever, or sell context filtering as the contribution.
The purpose is to test whether plausible but temporally stale repository
context can induce code that is incompatible with the current repository state.

\subsection{Research Questions}

RQ1: Does stale repository context induce current-state-incompatible helper
calls compared with current repository context and no retrieval?

RQ2: When both current and stale evidence are present, does rank order change
the stale-reference rate?

RQ3: Does adding valid current evidence to a prompt that contains stale evidence
reduce the stale-reference rate?

RQ4: Are stale-triggering samples shared across an open-source code-specialist
model and a proprietary general-purpose model?

\subsection{No-Target-Anchor Prompting Policy}

Earlier pilot gates in our study showed that strong models can self-rescue when the task
explicitly names the intended API, library, or repository state. Therefore this
study uses no-target-anchor prompting.

\begin{itemize}
\item The prompt does not reveal whether retrieved snippets are current or stale.
\item The task does not name the expected current helper signature.
\item Commit hashes and freshness labels are oracle-only metadata.
\item The model-visible evidence consists of a local call-site wrapper and the retrieved repository snippet.
\end{itemize}

This policy is enforced at the dataset-construction level: candidate samples
that require revealing the current state to be meaningful are excluded by the
mining and curation pipeline rather than weakening the prompting policy. The
runtime harness also performs a lightweight static check for obvious freshness
or target-reference leakage, but the main enforcement is sample construction
and review.

\subsection{Sample Mining}

We mine candidate temporal-validity examples from real Python repository
history. The final neutralized dataset uses five Python repositories: \texttt{click},
\texttt{flask}, \texttt{httpx}, \texttt{requests}, and \texttt{rich} (see
supplementary mining report).

The mining pipeline scans recent commit history for Python function signature
changes. Each candidate stores the parent commit as stale context and the child
commit as current context. Candidates are filtered to remove:

\begin{itemize}
\item test functions and files under test directories
\item zero-delta signatures with no static-oracle power
\item duplicate helper families across nearby commits
\item fragile dunder or high-arity cases when they fail the quality gate
\end{itemize}

Offline mining produced 31 candidates. Curation reduced these to 19 samples,
and the Qwen current-context quality gate retained 17 samples.

Because Qwen is both the quality-gate model and one of the evaluated models, we
also keep the pre-gate 19-sample set as a sensitivity check. This separates the
conditioned-on-model-capable estimate from the less filtered estimate.

The final neutralized evaluation set contains only production/helper samples:

\begin{center}
\begin{tabular}{lrrrrrr}
\toprule
Repository & \texttt{click} & \texttt{flask} & \texttt{httpx} & \texttt{requests} & \texttt{rich} & Total \\
\midrule
Samples & 3 & 3 & 3 & 3 & 5 & 17 \\
\bottomrule
\end{tabular}
\end{center}

The signature-delta distribution is 13 samples with a one-argument change,
two samples with a two-argument change, and two samples with a three-argument
change.

All retained samples are \texttt{signature\_change} examples with one-parent-commit drift.
Rename/remove changes and larger temporal gaps are left for follow-up work or a
future expansion.

\subsection{Retrieval Conditions}

For each sample, the harness constructs five retrieval conditions. The
\texttt{current\_context\_only} condition retrieves the current helper
signature from the child commit, while \texttt{stale\_context\_only} retrieves
the stale helper signature from the parent commit. The \texttt{no\_retrieval}
baseline supplies no cross-file repository snippet and tests whether stale
context induces wrong-state references rather than merely causing generic
model failure. The two mixed conditions include both snippets:
\texttt{mixed\_current\_top1\_stale\_top2} places the current snippet first, and
\texttt{mixed\_stale\_top1\_current\_top2} places the stale snippet first.

The primary comparison is stale-only versus current-only retrieval.
The \texttt{no\_retrieval} condition is a baseline. The mixed conditions are robustness
checks for rank-order sensitivity and current-evidence rescue, not the main
decision gate.

This is an oracle-controlled retrieval setting rather than a deployed-retriever
evaluation. The prompt intentionally asks the model to use retrieved repository
evidence because the diagnostic question is whether temporally invalid evidence
can induce wrong-state grounding when it is treated as project evidence. We
therefore interpret the stale-only condition as a controlled upper-bound
diagnostic of context-grounded failure, not as an estimate of stale-index
prevalence in deployed systems.

\subsection{Models}

The main evaluation uses two models:

\begin{center}
\begin{tabular}{ll}
\toprule
Model & Role \\
\midrule
\texttt{Qwen} & Open-source code-specialist model \\
\texttt{gpt-4.1-mini} & Proprietary general-purpose model \\
\bottomrule
\end{tabular}
\end{center}

Here \texttt{Qwen} abbreviates \texttt{Qwen2.5-Coder-7B-Instruct}.

DeepSeek-Coder-6.7B was explored during pilot work, but it is excluded from the
main analysis because its OpenAI-compatible serving path produced tokenizer
artifacts and low current-context pass rates under this harness.
This exclusion is documented in the supplementary run logs.

\subsection{Inference Configuration}

All model calls use the same OpenAI-compatible chat-completions client from the
experiment harness. The request payload explicitly sets \texttt{temperature=0};
the client implementation is included in the supplementary artifact.

The harness does not explicitly set \texttt{top\_p}, \texttt{seed}, or \texttt{max\_tokens}; these
remain provider or vLLM endpoint defaults. Qwen runs use the local vLLM
OpenAI-compatible endpoint, while \texttt{gpt-4.1-mini} runs use the OpenAI API
endpoint. Request delays affect rate limiting only and are not part of the
decoding configuration.

\subsection{Metrics}

The primary metric is stale-reference rate:

\[
\text{stale-reference rate} =
\frac{\#\{\text{outputs matching the stale helper call pattern}\}}{n}.
\]

The primary effect is:

\[
\Delta_{\text{primary}} =
\text{SRR}_{\text{stale-only}} - \text{SRR}_{\text{current-only}}.
\]

For the no-retrieval baseline, we report:

\[
\Delta_{\text{no-retrieval}} =
\text{SRR}_{\text{stale-only}} - \text{SRR}_{\text{no-retrieval}}.
\]

This separates stale-reference errors from generic inability to complete the
wrapper without repository context.

We also report pass rate:

\[
\text{pass rate} =
\frac{\#\{\text{outputs matching current patterns and not stale patterns}\}}{n}.
\]

This matters because a model can avoid stale references while still failing to
produce the current-state call. For example, \texttt{gpt-4.1-mini} has a 0/17
current-context stale-reference rate in the final neutralized evaluation, but its current-context pass rate is
15/17 rather than 17/17.

We separately report fail-no-match rate:

\[
\text{fail-no-match rate} =
\frac{\#\{\text{outputs matching neither oracle}\}}{n}.
\]

This captures cases where the model output does not satisfy either static
oracle. It is reported separately from stale-reference rate because it can
indicate oracle strictness or model-format drift rather than a stale-context
failure.

For mixed conditions, we report rank-order stale-reference delta:

\[
\Delta_{\text{rank}} =
\text{SRR}_{\text{stale-top1}} - \text{SRR}_{\text{current-top1}}.
\]

Finally, we compute cross-model sample overlap among stale-triggering samples:

\[
J =
\frac{|S_{\text{Qwen}} \cap S_{\text{GPT}}|}
{|S_{\text{Qwen}} \cup S_{\text{GPT}}|}.
\]

\subsection{Oracles}

The final neutralized evaluation uses static call-pattern oracles. For each sample, the
builder generates:

\begin{itemize}
\item current call regex patterns from the child-commit signature
\item stale call regex patterns from the parent-commit signature
\item expected current solution sketch
\item forbidden stale reference sketch
\end{itemize}

The oracle is intentionally conservative. Outputs that do not match either the
current or stale pattern are counted as missing current signal, not as stale
references.

\subsection{Prompt-Leakage Audit and Neutralized Final Run}
\label{sec:prompt-audit}

Before the final evaluation, we performed a prompt-leakage audit over every
model-visible prompt field: local context, retrieved-context headers,
context identifiers, task text, and optional metadata fields. The
audit found that preliminary rendered prompts used condition-specific
context identifiers that could encode freshness cues. Because this violated
our no-target-anchor policy, we excluded the preliminary run from the main
analysis.

We then constructed a neutralized evaluation set in which all model-visible
context identifiers were replaced with opaque per-sample identifiers
(e.g., \texttt{ctx-sig-001-a} and \texttt{ctx-sig-001-b}), assigned by a
deterministic hash so that no positional convention across samples can be
learned. In the retained 17-sample set, the \texttt{a} identifier holds
the current snippet for 9 samples and the \texttt{b} identifier holds it
for the other 8. The complete five-condition protocol was rerun for both models. All
results reported below come from this neutralized run. The excluded preliminary
run and the audit history are provided in Appendix~\ref{app:prompt-audit} for
transparency and are not used as evidence for the main claims.

\subsection{Validity-Threat Summary}

The design choices above make the treatment controlled but narrow.
Section~\ref{sec:limitations} discusses sample size, signature-change scope,
one-parent drift, oracle-controlled retrieval, static oracles, Qwen
quality-gate entanglement, backend reproducibility, and the prompt-leakage
audit disclosure. We keep the detailed threat discussion there to avoid mixing
design definition with validity analysis.

\section{Results}

\subsection{RQ1: Stale Context Strongly Increases Stale References}

On the 17-sample neutralized evaluation set, both main models show large
stale-reference deltas:

\begin{center}
\small
\begin{tabular}{llll}
\toprule
Model & Current condition & Stale condition & $\Delta_{\text{primary}}$ \\
\midrule
\texttt{Qwen} & SRR 0/17; pass 17/17 & SRR 15/17; pass 0/17 & 88.2 pp \\
\texttt{gpt-4.1-mini} & SRR 0/17; pass 15/17 & SRR 13/17; pass 1/17 & 76.5 pp \\
\bottomrule
\end{tabular}
\end{center}

Here \texttt{Qwen} abbreviates \texttt{Qwen2.5-Coder-7B-Instruct}.

Interpretation:

\begin{quote}\emph{When retrieved repository context is the model-visible source of repo-state evidence, stale helper signatures can induce current-state-incompatible calls in both \texttt{Qwen2.5-Coder-7B-Instruct} and \texttt{gpt-4.1-mini}.}\end{quote}

We report stale-reference delta rather than only pass-rate delta because some
outputs fail to match either the current or stale static pattern. The
\texttt{gpt-4.1-mini} current-context condition has two such fail-no-match cases
(\texttt{sig-013}, \texttt{sig-026}), so it must not be described as 17/17 current-pass.

\subsubsection{Failure Mode Taxonomy}

To make the aggregate stale-reference rate inspectable, we manually categorized
the stale-triggering outputs from the two main \texttt{stale\_context\_only} runs. This
analysis covers 28 stale-positive output events: 15 from \texttt{Qwen2.5-Coder-7B-Instruct} and
13 from \texttt{gpt-4.1-mini}. These events correspond to 16 unique samples because
the two models agree on 12 stale-triggering samples, Qwen contributes three
model-specific stale triggers, and \texttt{gpt-4.1-mini} contributes one.

\begin{table}[H]
\centering
\scriptsize
\caption{Failure-mode taxonomy for 28 stale-positive output events across the two main \texttt{stale\_context\_only} runs. The examples show representative sample IDs and the stale call shape.}
\label{tab:failure-taxonomy}
\begin{tabular}{p{0.20\linewidth}p{0.12\linewidth}p{0.28\linewidth}p{0.28\linewidth}}
\toprule
Failure mode & Count & Typical shape & Example \\
\midrule
New argument omitted & 22/28 & The current helper signature requires one or more additional arguments, but the model calls the stale shorter signature. & \texttt{sig-027}: stale two-argument \texttt{chop\_cells} call instead of the current three-argument call. \\
Obsolete argument retained & 5/28 & The stale helper signature contains an argument no longer used by the current helper, and the model keeps it. & \texttt{sig-001}: stale three-argument \texttt{\_nullpager} call instead of the current two-argument call. \\
Stale helper reconstruction & 1/28 & The model reconstructs a helper body or wrapper around the stale signature rather than simply making a stale call. & \texttt{sig-003}: Qwen defines a local stale-shape \texttt{\_tempfilepager} helper before using it. \\
\bottomrule
\end{tabular}
\end{table}

The dominant error is therefore not an arbitrary syntax failure. In most
stale-positive outputs, the model produces a plausible call whose argument list
matches the retrieved historical helper rather than the current repository
state. This strengthens the interpretation that stale context acts as a
directional misleading signal: it changes the shape of the generated call
toward a concrete historical signature.

The taxonomy also explains why pass rate and stale-reference rate should remain
separate. Some outputs match neither current nor stale patterns, especially in
high-arity wrapper samples. Those fail-no-match cases are not counted in this
taxonomy because they do not provide evidence that the model followed stale
repository state.

\subsubsection{No-Retrieval Baseline}

Both models show the same no-retrieval baseline:

\begin{center}
\small
\begin{tabular}{lrrr}
\toprule
Model & No-retrieval SRR & No-retrieval pass & Fail-no-match \\
\midrule
\texttt{Qwen} & 0/17 (0.0\%) & 1/17 (5.9\%) & 16/17 (94.1\%) \\
\texttt{gpt-4.1-mini} & 0/17 (0.0\%) & 1/17 (5.9\%) & 16/17 (94.1\%) \\
\bottomrule
\end{tabular}
\end{center}

Compared with no retrieval, stale-only retrieval does not merely cause generic
failure. It specifically shifts many outputs toward stale helper references:

\[
\Delta_{\text{stale vs. no-retrieval}} =
88.2\% - 0.0\% = 88.2 \text{ percentage points (Qwen)}
\]
\[
\Delta_{\text{stale vs. no-retrieval}} =
76.5\% - 0.0\% = 76.5 \text{ percentage points (\texttt{gpt-4.1-mini})}.
\]

The matching no-retrieval result matters because it separates two failure
types. Without retrieval, both models mostly produce outputs that match neither
oracle. With stale retrieval, the dominant failure becomes stale-state helper
references. Thus stale context is actively misleading rather than merely
uninformative.

The only no-retrieval pass in both models is \texttt{sig-023}
(\texttt{requests Session.get}), a common API likely recoverable from
parametric model knowledge.

\subsubsection{Sensitivity: Qwen Without Quality-Gate Filtering}

The main Qwen table uses 17 samples retained by a Qwen current-context quality
gate. To quantify this entanglement, we also ran Qwen on the 19 curated samples
before that quality gate:

\begin{center}
\small
\begin{tabular}{lrrr}
\toprule
Set & Current pass & Current SRR & Stale SRR \\
\midrule
Qwen n=19 sensitivity & 18/19 (94.7\%) & 0/19 (0.0\%) & 15/19 (78.9\%) \\
\bottomrule
\end{tabular}
\end{center}

The corresponding primary delta is 78.9 percentage points.

The gate-conditioned n=17 estimate is therefore 9.3 percentage points higher than the
ungated n=19 estimate. Both estimates remain far above our pilot continuation
threshold of 20 percentage points, so the main effect does not depend on
hiding the quality-gate entanglement. The n=19 sensitivity run was executed
after the original quality-gate decision. In that rerun, \texttt{sig-012}
passed current-context retrieval, while \texttt{sig-024} still failed. We
retain the original n=17 main set to avoid post-hoc redefinition of the quality
gate and report n=19 only as sensitivity evidence.

\subsection{RQ2: Rank Order Does Not Consistently Amplify the Failure}

Qwen mixed-condition results:

\begin{center}
\small
\begin{tabular}{lrl}
\toprule
Mixed order & Stale-reference rate & Stale-reference samples \\
\midrule
Current first & 4/17 (23.5\%) & \texttt{sig-001}, \texttt{sig-002}, \texttt{sig-008}, \texttt{sig-010} \\
Stale first & 4/17 (23.5\%) & \texttt{sig-001}, \texttt{sig-007}, \texttt{sig-010}, \texttt{sig-013} \\
\bottomrule
\end{tabular}
\end{center}

The Qwen rank-order delta is 0.0 percentage points. The two orders have the
same aggregate rate, with two shared stale-triggering samples and two
order-specific samples on each side.

\texttt{gpt-4.1-mini} mixed-condition results:

\begin{center}
\small
\begin{tabular}{lr}
\toprule
Mixed order & Stale-reference rate \\
\midrule
Current first & 5/17 (29.4\%) \\
Stale first & 5/17 (29.4\%) \\
\bottomrule
\end{tabular}
\end{center}

For \texttt{gpt-4.1-mini}, the rank-order delta is 0.0 percentage points; the
same five samples trigger under both mixed orders.

Interpretation:

\begin{quote}\emph{We do not find evidence that placing stale context at rank 1 amplifies the failure in the neutralized evaluation. Both models have 0.0 percentage-point aggregate rank-order deltas. Qwen's equality comes from compensating order-specific samples, while \texttt{gpt-4.1-mini}'s equality comes from identical sample-level triggers across both orders.}\end{quote}

This result is consistent with our earlier pilot observations on version-skew
and cross-library distractors. We therefore treat it as evidence under the
current task framing rather than as a universal claim that rank order never
matters.

\subsection{RQ3: Current Evidence Presence Rescues Most Stale-Only Failures}

Adding valid current evidence reduces stale references from 88.2\% (Qwen) and
76.5\% (\texttt{gpt-4.1-mini}) under \texttt{stale\_context\_only} to much lower
rates under mixed conditions.

For Qwen:

\[
\text{presence-absence rescue}_{\text{Qwen}} =
88.2\% - 23.5\% = 64.7 \text{ percentage points}.
\]

For \texttt{gpt-4.1-mini}, both mixed orders produce the same stale-reference
rate and remain well below the stale-only condition:

\[
\begin{aligned}
\text{rescue}_{\texttt{gpt-4.1-mini}}
&= 76.5\% - 29.4\% \\
&= 47.1\text{ percentage points}.
\end{aligned}
\]

This suggests that the dominant failure mechanism is not rank order itself, but
whether the valid current evidence is retrieved at all.

\subsection{RQ4: Qwen and GPT-Mini Have Substantial Sample-Level Agreement}

Qwen produces 15 stale-triggering samples and \texttt{gpt-4.1-mini} produces 13
under \texttt{stale\_context\_only}.
Their overlap is:

\begin{center}
\small
\begin{tabular}{lr}
\toprule
Quantity & Value \\
\midrule
Qwen stale-triggering samples & 15 \\
\texttt{gpt-4.1-mini} stale-triggering samples & 13 \\
Intersection & 12 \\
Union & 16 \\
Jaccard overlap & 75.0\% \\
Wilson 95\% interval for Jaccard & [50.5\%, 89.8\%] \\
Qwen coverage by \texttt{gpt-4.1-mini} & 80.0\% \\
\texttt{gpt-4.1-mini} coverage by Qwen & 92.3\% \\
\bottomrule
\end{tabular}
\end{center}

The overlap breakdown is 12 samples triggered by both models, three Qwen-only
samples (\texttt{sig-016}, \texttt{sig-023}, \texttt{sig-026}), one
\texttt{gpt-4.1-mini}-only sample (\texttt{sig-028}), and one neither-trigger sample
(\texttt{sig-013}).

Per-repository agreement:

\begin{table}[H]
\centering
\small
\caption{Sample-level agreement by repository under \texttt{stale\_context\_only}. The table reports whether stale-triggering samples are shared by both models or model-specific.}
\label{tab:model-agreement}
\begin{tabular}{lrrrrr}
\toprule
Repo & n & Both trigger & Qwen-only & GPT-only & Neither \\
\midrule
click & 3 & 3 & 0 & 0 & 0 \\
flask & 3 & 3 & 0 & 0 & 0 \\
requests & 3 & 2 & 1 & 0 & 0 \\
rich & 5 & 3 & 1 & 1 & 0 \\
httpx & 3 & 1 & 1 & 0 & 1 \\
Total & 17 & 12 & 3 & 1 & 1 \\
\bottomrule
\end{tabular}
\end{table}

Interpretation:

\begin{quote}\emph{On the neutralized dataset, the stale-context failure mode is not merely an aggregate artifact: Qwen and \texttt{gpt-4.1-mini} agree on most stale-triggering samples.}\end{quote}

The confidence interval is wide because the union contains only 16 positive
samples. We therefore use Jaccard overlap as supporting evidence of
sample-level agreement, not as a precise estimate of cross-model stability.

However, we should not attribute the improvement over earlier Qwen/DeepSeek
pilot overlap solely to curation, because both the dataset and model pair
changed.

Note: sample \texttt{sig-013}, an \texttt{httpx} transport helper, has no
stale-reference classification under stale-only retrieval for either model. We
treat this as a possible oracle-strictness case rather than strong evidence of
model robustness: the output may use a keyword-shifted or receiver-prefixed
call form that the current static regex does not capture. A future iteration
should review this sample with relaxed oracle patterns before re-classifying it.

\subsection{Excluded Samples}

The Qwen current-context quality gate started from 19 curated samples and
retained 17:

\texttt{sig-012} (\texttt{httpx Client.\_init\_proxy\_transport}) and
\texttt{sig-024} (\texttt{rich render\_scope}) were excluded because Qwen failed
the original current-context quality gate on these high-arity wrapper tasks.

These are not counted in the main table. They should be reported as a
limitation of the current task-wrapper format rather than as model failures
under stale retrieval.

The n=19 sensitivity run includes these two rows and still produces a 78.9 percentage-point
Qwen primary delta. We treat this as a sensitivity check for the quality-gate
entanglement, not as an alternative main table. In the neutralized
sensitivity rerun, \texttt{sig-012} passed current-context retrieval while
\texttt{sig-024} remained fail-no-match. We keep the original gate decision for
protocol consistency rather than reclassifying the main set after inspecting a
later rerun.

\subsection{Statistical Note}

For Qwen's 15/17 stale-reference rate, the Wilson 95\% interval is
approximately 65.7\% to 96.7\%; for \texttt{gpt-4.1-mini}'s 13/17 rate, it is
approximately 52.7\% to 90.4\%. These Wilson intervals describe stale-only
rates rather than the paired treatment effect, and they remain too wide for
tight population-level effect-size estimation. For the cross-model Jaccard
overlap, 12/16 gives an approximate Wilson interval of 50.5\% to 89.8\%.

Paired McNemar exact two-sided tests on the primary contrast
(\texttt{stale\_context\_only} versus \texttt{current\_context\_only}) yield
$p = 6.10 \times 10^{-5}$ for \texttt{Qwen2.5-Coder-7B-Instruct} and
$p = 2.44 \times 10^{-4}$ for \texttt{gpt-4.1-mini}, paired within sample.
For rank order, the exact paired tests yield $p = 1.0$ for both models. In
\texttt{gpt-4.1-mini}, the two mixed orders produce no discordant
stale-reference samples, so the test should be read as no observed
sample-level rank-order difference in this run rather than as proof that
rank can never matter. Rescue contrasts between stale-only and
mixed conditions are also significant for both models ($p \leq 7.81 \times
10^{-3}$).

The paired tests strongly support the existence of the stale-context effect in
this curated diagnostic set. However, the sample is still too small for precise
population-level effect-size estimation or detailed subgroup analysis.

\subsection{Takeaway}

The evidence supports a controlled diagnostic study with three main results:

\begin{enumerate}
\item Stale-only context induces large stale-reference deltas in both Qwen and \texttt{gpt-4.1-mini}.
\item Adding valid current evidence largely rescues the failure; rank order does not consistently amplify it.
\item The two models agree on most stale-triggering samples in the curated set, though sample size and axis coverage remain explicit limitations.
\end{enumerate}

\section{Discussion}

\subsection{Stale Retrieval Is Not the Same as Missing Retrieval}

The no-retrieval baseline changes how the main result should be interpreted.
Without retrieved repository context, both Qwen and \texttt{gpt-4.1-mini}
rarely produce the expected current helper call: each model
passes only 1 of 17 samples and otherwise produces outputs that match neither
the current nor stale oracle. However, neither model emits stale references
under no retrieval. In contrast, stale-only retrieval induces stale references
on 15 of 17 Qwen samples and 13 of 17 \texttt{gpt-4.1-mini} samples.

This distinction matters. A stale snippet is not merely absent information or
generic retrieval noise. It provides a concrete but outdated state hypothesis:
the model can use it to produce syntactically plausible code that is anchored
to an obsolete helper signature. In our setting, stale retrieval changes the
error pattern from fail-no-match to wrong-state grounding. This is the core
diagnostic value of the experiment.

The condition contrast also argues against a simpler explanation: that the
models merely copy whatever retrieved snippet they see. Current-only retrieval
also supplies a concrete helper snippet, but the stale-reference rate remains
0.0\% under that condition. The failure appears when the retrieved helper is
temporally invalid, not whenever retrieval text is present. This suggests that
the model is grounding on the repository state implied by the retrieved
snippet, rather than performing content-agnostic copying from retrieval.

\subsection{Valid Evidence Presence Dominates Rank Order}

The mixed conditions show that current evidence can rescue most stale-only
failures. For Qwen, adding current context reduces stale-reference rate from
88.2\% under stale-only retrieval to 23.5\% under mixed retrieval. For
\texttt{gpt-4.1-mini}, both mixed orders produce 29.4\% stale-reference rates,
far below stale-only retrieval. This suggests that the dominant question is
whether valid current evidence is retrieved at all.

Rank order appears secondary in the current sample. Both models have 0.0
percentage-point aggregate rank-order deltas in the neutralized evaluation.
The sample-level behavior is still informative: Qwen's equal rates come from
two shared mixed-condition stale triggers and compensating order-specific
triggers, whereas \texttt{gpt-4.1-mini}'s equal rates come from the same five
samples triggering under both orders. A larger study could test whether
different models have weak rank-order preferences once both current and stale
evidence are present, but our evidence points to current-evidence recall as
the stronger factor.

\subsection{Why No-Target-Anchor Prompting Matters}

The result also explains why two earlier pilot ideas failed. In the
version-skew and cross-library pilots, the task often explicitly named the
target API, library, or intended state. Under those prompts, strong models
treated retrieved distractors as advisory and followed the explicit target
anchor. In the current stale-repo setup, the prompt removes that anchor: the
retrieved snippets are the model-visible source of repository-state evidence.

This suggests a methodological implication for future Code RAG robustness studies.
If the research question is whether retrieval can induce a wrong state, the
task prompt should not independently reveal the correct state. Otherwise the
experiment may measure the model's ability to ignore retrieval under strong
instructional anchoring rather than the model's reliance on retrieved evidence.

\subsection{Implications for Code RAG Systems}

The study does not propose a freshness filter or retriever. Still, the results
suggest where such systems should be careful. Repository-level retrieval
pipelines should treat temporal validity as a first-class metadata property,
not merely as a side effect of relevance. A retrieved snippet can be highly
topical, from the same project, and close to the target helper name while still
being harmful because it reflects an older repository state.

This is especially relevant for systems that cache indexes, retrieve from code
search snapshots, use external documentation mirrors, or combine local context
with historical examples. If the current state is not retrieved, the model may
not simply abstain or fail generically; it may confidently bind to the stale
state. The no-retrieval comparison suggests that stale evidence can be worse
than no evidence along the specific dimension of wrong-state references.

Concrete engineering checks follow naturally from this result. Retrieval
systems can attach commit-hash or index-build-time freshness headers to
retrieved snippets, enforce index-invalidation windows when the repository
changes, or run a current-evidence recall probe before generation. These checks
would need their own evaluation; we list them as implications rather than as
claims of a mitigation contribution.

\subsection{What This Paper Does Not Claim}

We do not claim that all harmful retrieval is unexplored, that context
filtering is new, or that rank order never matters. Prior work already studies
harmful retrieved chunks, selective retrieval, dependency-aware mitigation, and
repository-level code completion. Our claim is narrower: to our knowledge, the
closest related work reviewed here does not control old-vs-current retrieved
repository context as the treatment variable and report stale-reference
behavior under matched completion tasks.

We also do not claim a deployed mitigation. The current contribution is a
diagnostic protocol and a set of empirical observations. A natural next step is
to test whether simple freshness metadata, commit-aware index invalidation, or
current-evidence recall checks reduce stale-reference failures. Those systems
should be evaluated as future mitigation work rather than folded into the
present diagnostic study.

\section{Limitations}
\label{sec:limitations}

\subsection{Sample Size}

The main evaluation uses 17 retained production-helper samples. The stale-only
effect is large enough to motivate a controlled diagnostic study, but the sample is
not large enough for tight effect-size estimation or detailed subgroup
analysis. The Wilson intervals for the 15/17 Qwen and 13/17
\texttt{gpt-4.1-mini} stale-reference rates remain wide, and the 75.0\%
cross-model Jaccard overlap also has a wide interval. We therefore treat the
estimates as diagnostic evidence rather than final population-level rates.

The n=19 Qwen sensitivity run partially addresses selection concerns: before
Qwen current-context quality gating, the stale-reference delta is 78.9
percentage points rather than the gate-conditioned 88.2 percentage points.
The effect survives the ungated curated set, but this does not
replace a larger evaluation. A future expansion should move toward 40--60
samples if the goal is tighter confidence intervals or per-repository analysis.

\subsection{Prompt-Leakage Audit Disclosure}

A preliminary version of the evaluation used model-visible context identifiers
that encoded condition-specific terms. This violated our no-target-anchor
policy and was caught during a pre-submission prompt-leakage audit. We
therefore neutralized all model-visible context identifiers and reran the full
evaluation. All main results in this paper use the neutralized run. This
experience suggests that Code RAG diagnostic evaluations should audit not just
task text and retrieved snippets, but also seemingly administrative fields such
as context identifiers and metadata headers.

\subsection{Axis Coverage}

All retained samples are signature-change examples with one-parent-commit drift. This
keeps the first study controlled, but it leaves out other important temporal
change types: renamed helpers, removed helpers, moved modules, refactors that
preserve signatures but change semantics, and cross-release drift. The current
study therefore should not be generalized to all forms of stale repository
context. It shows that stale retrieval can induce wrong-state calls for signature drift;
future work should test whether the same mechanism holds across broader change
types and longer temporal gaps.

\subsection{Repository and Language Scope}

The dataset uses five Python repositories: \texttt{click}, \texttt{flask}, \texttt{httpx},
\texttt{requests}, and \texttt{rich}. These projects provide realistic production-helper
changes, but they do not cover other languages, typed build systems, monorepo
structures, generated code, or compiled dependency boundaries. The result is
most directly about Python repository-helper completion under static
call-pattern oracles. Applying the protocol to Java, Rust, TypeScript, or
multi-language repositories may reveal different failure rates and different
oracle requirements.

The repository selection is also biased toward mature public libraries with
clean histories and reviewable helper functions. Private monorepos, generated
code, rapidly changing product repositories, or codebases with weaker tests may
exhibit different retrieval-freshness risks. The current sample should
therefore be read as evidence that the failure mode exists in realistic
open-source libraries, not as a population estimate for all repositories.

\subsection{Task Framing}

The task is a local wrapper completion around a single helper call. This keeps
the treatment controlled and makes stale-reference behavior inspectable, but it
may increase reliance on the retrieved helper snippet compared with a richer
IDE session containing more surrounding files, tests, or developer intent. A
larger follow-up should test whether the effect persists for broader edit
tasks and multi-function completions.

The no-target-anchor prompting policy is also load-bearing. It is appropriate
for testing whether retrieved context can determine the model-visible
repository state, but it should not be generalized to prompts that explicitly
name the intended current API or commit state. Earlier pilots suggest that such
anchors can let strong models ignore distractors, so both prompt families are
important but answer different questions.

\subsection{Static Oracles}

The current evaluation uses static regex-style call-pattern oracles rather than
full executable tests. This makes the experiment cheap and repeatable, but it
can miss semantically equivalent variants. Sample \texttt{sig-013} is the clearest
case: both models avoid a stale-reference classification under stale-only
retrieval, but the output may use a keyword-shifted, receiver-prefixed, or
otherwise valid form that the static oracle does not capture. We therefore
report fail-no-match separately from stale-reference rate and avoid treating
every fail-no-match as model robustness.

Executable tests would strengthen the evidence. They would also introduce
engineering costs: dependency reconstruction, historical environment setup,
and per-commit test harnesses. The present study prioritizes a controlled
diagnostic signal; a follow-up should convert the strongest samples into
executable tests.

\subsection{Quality-Gate Entanglement}

The 17-sample main set is filtered by Qwen current-context success. This makes
Qwen's 17/17 current pass rate a construction property of the retained sample
set rather than an independent finding. We address this in two ways. First, we
report \texttt{gpt-4.1-mini} current-context pass separately; it was not used for the
quality gate and passes 15/17 samples. Second, we run Qwen on the 19 curated
samples before quality-gate filtering, where the stale-reference delta remains
78.9 percentage points. This quantifies the Qwen-gate shift at about 9.3 percentage points, but it does not
eliminate the selection issue entirely.

\subsection{Model Coverage}

The main evidence uses two models: \texttt{Qwen2.5-Coder-7B-Instruct} and
\texttt{gpt-4.1-mini}. This is enough to show that the effect is not limited to a
single model, but it is not a broad model-family study. DeepSeek-Coder-6.7B was
explored during pilot work, but its OpenAI-compatible serving path produced
tokenizer artifacts and low current-context pass rates under this harness. We
exclude it from the main analysis rather than forcing noisy results into the
paper.

The mixed-condition results also suggest that aggregate rank-order invariance
can arise from different sample-level patterns. Qwen has the same aggregate
mixed-condition stale-reference rate under both orders but different
order-specific samples, while \texttt{gpt-4.1-mini} triggers on the same five
samples under both orders. Larger samples and additional models are needed
before making claims about rank-order sensitivity across model families.

\subsection{Generation Variability}

Most runs were stable at the aggregate level, but at least one edge sample
changed classification across repeated Qwen current-context runs. The harness
explicitly sends \texttt{temperature=0} in the OpenAI-compatible request payload, so
this is not intended sampling diversity. However, the result rows do not store
the full request body or the serving backend's effective sampling state. We
therefore treat the edge-case variation as backend or output-format variability
and note that future runs should log backend versions, provider defaults,
effective sampling parameters, and seed support explicitly.

\subsection{Dataset Independence}

Samples are mined from upstream repository history rather than derived from
nearby published datasets. This avoids directly reusing benchmark artifacts from
nearby work, but it also means our dataset is smaller and more labor-intensive
to construct. We make the sample construction process
transparent enough that future work can extend it without relying on our exact
sample set.

\section{Reproducibility Statement}

The experimental harness, sample files, oracle patterns, and raw outputs are
stored under \path{experiments/stale_repo_rag_pilot/}. The final neutralized
evaluation set is \path{data/curated_v3_neutralized.jsonl} (17 samples), with
the pre-quality-gate sensitivity set in
\path{data/curated_candidates_v3_neutralized.jsonl} (19 samples). Sample
construction rules are documented in the mining and curation artifacts under
\path{experiments/stale_repo_rag_pilot/runs/}. The neutralization script,
prompt validator, prompt dumps, and mapping file are released with the
artifact so that readers can verify that model-visible context identifiers do
not encode freshness labels.

All model calls use the OpenAI-compatible chat-completions client in
\path{stale_repo_rag_pilot/clients.py}. The request payload explicitly sets
\texttt{temperature=0}; \texttt{top\_p}, \texttt{seed}, and \texttt{max\_tokens} are left to provider or
vLLM endpoint defaults. \texttt{Qwen2.5-Coder-7B-Instruct} is served through a local
vLLM OpenAI-compatible endpoint. \texttt{gpt-4.1-mini} is served through the OpenAI
API with the same prompt and oracle pipeline.

The raw JSONL result rows preserve generated outputs, oracle decisions, model
names, conditions, and rendered-prompt-derived classifications. They do not
reconstruct provider-internal sampling state beyond metadata exposed by the
OpenAI-compatible endpoints. We therefore report the client-side request
configuration from the harness and distinguish it from server-side effective
sampling state, which may depend on provider or vLLM endpoint defaults.

The oracle implementation is in \texttt{stale\_repo\_rag\_pilot/oracle.py}. It strips
Markdown code fences, normalizes byte-level tokenizer artifacts observed in
pilot serving, and evaluates generated code with static current and stale call
patterns stored in each JSONL sample. Raw per-condition logs for the final
neutralized evaluation are preserved in:
\begin{itemize}
\item \path{runs/qwen-v3-primary/} and \path{runs/openai-v3-primary/}
\item \path{runs/qwen-v3-sensitivity-n19/} for the ungated Qwen sensitivity check
\item \path{runs/v3_smoke_prompt_dump/} for rendered prompt leakage checks
\end{itemize}
The preliminary run predating the prompt-leakage audit is retained only as an
excluded audit-history artifact and is not used to reproduce the main tables.
Citation metadata was batch-checked in the citation-verification artifact.

A self-contained artifact accompanies this submission as supplementary
material. It contains: the final neutralized 17-sample dataset and the
19-sample ungated sensitivity set; the per-sample opaque context-identifier
mapping file; V3 result logs for both models; the experimental harness
source code; the neutralization and prompt-audit scripts; and five
representative rendered prompts illustrating the model-visible inputs. An
automated leakage-audit summary over all 85 rendered prompts is also
included. The exact vLLM server version and model snapshot hashes used for
the released runs will be recorded in the camera-ready artifact alongside a
permanent DOI.

\section{Conclusion}

This paper presents a controlled diagnostic study of stale repository context in
retrieval-augmented code completion. Across 17 production-helper signature
changes from five Python repositories, stale-only retrieved context induces
stale-state helper references on 15 of 17 Qwen samples and 13 of 17
\texttt{gpt-4.1-mini} samples, while current-only context eliminates stale references. The
no-retrieval baseline shows a different failure mode: without repository
context, both models mostly fail to match either oracle rather than producing
stale references. This distinction suggests that stale retrieval is actively
misleading, not merely unhelpful.

The results also clarify the mechanism. When valid current evidence is present
alongside stale evidence, stale-reference rates drop sharply. Both models show
0.0 percentage-point aggregate rank-order deltas across mixed conditions. The
dominant factor in this
study is therefore whether valid current evidence is retrieved at all, not the
relative ordering of stale and current snippets once both are available.

These findings point to a concrete future expansion path. Larger studies should expand to
40-60 samples across rename, remove, move, and semantic-refactor axes; add
executable oracles for stronger validation; and test additional model families
and languages. System builders should treat temporal validity as a first-class
retrieval property: highly relevant repository context can still be harmful if
it represents an obsolete project state.

\appendix

\section{Prompt Templates}

\path{experiments/stale_repo_rag_pilot/stale_repo_rag_pilot/prompts.py}.

\subsection{System Prompt}

\begin{verbatim}
You are a coding assistant. Return only code, no explanation.
Use the local file context and retrieved repository context
as the available project evidence.
\end{verbatim}

\subsection{User Prompt Template}

\begingroup\footnotesize
\begin{verbatim}
Local file context:
{model_visible_local_context}

Task:
{task_prompt}

Retrieved repository context:
{retrieved_context_blocks}

Return only a minimal {language} code snippet that completes the task.
\end{verbatim}
\endgroup

The model-visible local context is filtered before prompt construction. Commit
hashes, freshness labels, target-state metadata, expected current references,
and forbidden stale references are oracle-only and are not included in the
prompt. Retrieved context blocks expose an internal context id and the snippet
text. File paths can be exposed by an optional \texttt{--context-metadata path} flag,
but the final neutralized runs use hidden context metadata.

\subsection{Worked Example: \texttt{stale-repo-static-sig-001}}

The following is the verbatim \texttt{stale\_context\_only} prompt shape for a
Click helper sample from the neutralized run. The context id is model-visible
but opaque; the mapping from this id to freshness state is oracle-only and not
present in the prompt.

\begingroup\footnotesize
\begin{verbatim}
Local file context:
{
  "file_path": "generated/call_site.py",
  "prefix": "def invoke_target(stream, color, generator):\n    "
}
Task:
Complete invoke_target(stream, color, generator)
by calling the repository helper shown in the retrieved context.

Retrieved repository context:
[CONTEXT id=ctx-sig-001-a]
def _nullpager(
    stream: t.TextIO, generator: cabc.Iterable[str], color: bool | None
) -> None:
    """Simply print unformatted text.  This is the ultimate fallback."""
    for text in generator:
        if not color:
            text = strip_ansi(text)
        stream.write(text)

Return only a minimal python code snippet that completes the task.
\end{verbatim}
\endgroup

For the corresponding \texttt{current\_context\_only} condition, the retrieved context
contains \texttt{[CONTEXT id=ctx-sig-001-b]} with the current
\texttt{\_nullpager(stream, color)} signature instead. For the
\texttt{mixed\_*} conditions, both snippets are included and only their order
changes.

\subsection{No-Target-Anchor Guard}

The dataset construction policy forbids model-visible hints such as "current
HEAD", "stale context", "old commit", "target commit", and explicit current or
stale helper references in the local task. The runtime harness performs a
lightweight static check for these leakage terms before issuing model calls.

\section{Excluded Preliminary Run}
\label{app:prompt-audit}

The preliminary evaluation predating the prompt-leakage audit is reported here
only to document the audit history. It is \emph{excluded} from the main
evidence because its model-visible context identifiers contained
condition-specific terms.

\begin{table}[H]
\centering
\small
\caption{Aggregate comparison between the excluded preliminary run and the
neutralized final run. The preliminary run is not used as evidence for any
claim in this paper.}
\label{tab:excluded-preliminary}
\scriptsize
\begin{tabular}{p{0.40\linewidth}p{0.25\linewidth}p{0.22\linewidth}}
\toprule
Metric & Excluded preliminary run & Final neutralized run \\
\midrule
Qwen stale-only SRR & 14/17 & 15/17 \\
GPT-mini stale-only SRR & 14/17 & 13/17 \\
Aggregate stale-only events & 28/34 & 28/34 \\
Qwen rank-order $\Delta$ & 0.0 pp & 0.0 pp \\
GPT-mini rank-order $\Delta$ & $-17.6$ pp & 0.0 pp \\
Cross-model Jaccard & 75.0\% & 75.0\% \\
\bottomrule
\end{tabular}
\end{table}

The preliminary run is retained in the artifact only so that the audit trail is
interpretable. All main-paper tables and claims use the final neutralized run.

\bibliographystyle{elsarticle-num}
\bibliography{refs}

\end{document}